\documentclass[article, prX]{revtex4}
\usepackage{graphicx}
\begin{document}
\title{Imaging and mapping the impact of clouds on skyglow with all-sky photometry}

\author{Andreas Jechow}
\affiliation{Ecohydrology, Leibniz Institute of Freshwater Ecology and Inland Fisheries, M{\"u}ggelseedamm 310, 12587 Berlin}
\affiliation{Remote Sensing, Helmholtz Center Potsdam, German Center for Geosciences GFZ, Telegraphenberg, Potsdam}
\author{Zolt{\'a}n Koll{\'a}th}
\affiliation{Eötvös Loránd University, Savaria Department of Physics, Károlyi Gáspár tér 4, 9700 Szombathely, Hungary}
\author{Salvador J. Ribas}
\affiliation{Parc Astronòmic Montsec, Comarcal de la Noguera, Pg. Angel Guimerà 28-30, 25600 Balaguer, Lleida, Spain}
\affiliation{Institut de Ciències del Cosmos (ICCUB), Universitat de Barcelona, C.Martí i Franqués 1, 08028 Barcelona, Spain}
\author{Henk Spoelstra}
\affiliation{LightPollutionMonitoring.Net, Urb. Ve\"{i}nat Verneda 101 (Bustia 49), 17244 Cass\`{a} de la Selva, Girona, Spain}
\author{Franz H{\"o}lker}
\affiliation{Ecohydrology, Leibniz Institute of Freshwater Ecology and Inland Fisheries, M{\"u}ggelseedamm 310, 12587 Berlin}
\author{Christopher C. M. Kyba}
\affiliation{Remote Sensing, Helmholtz Center Potsdam, German Center for Geosciences GFZ, Telegraphenberg, Potsdam}
\affiliation{Ecohydrology, Leibniz Institute of Freshwater Ecology and Inland Fisheries, M{\"u}ggelseedamm 310, 12587 Berlin}

\maketitle
\section*{Abstract}

Artificial skyglow is constantly growing on a global scale, with potential ecological consequences ranging up to affecting biodiversity. To understand these consequences, worldwide mapping of skyglow for all weather conditions is urgently required. In particular, the amplification of skyglow by clouds needs to be studied, as clouds can extend the reach of skyglow into remote areas not affected by light pollution on clear nights. Here we use commercial digital single lens reflex cameras with fisheye lenses for all-sky photometry. We track the reach of skyglow from a peri-urban into a remote area on a clear and a partly cloudy night by performing transects from the Spanish town of Balaguer towards Montsec Astronomical Park. From one single all-sky image, we extract zenith luminance, horizontal and scalar illuminance.  While zenith luminance reaches near-natural levels at 5km distance from the town on the clear night, similar levels are only reached at 27km on the partly cloudy night. Our results show the dramatic increase of the reach of skyglow even for moderate cloud coverage at this site. The powerful and easy-to-use method promises to be widely applicable for studies of ecological light pollution on a global scale also by non-specialists in photometry.

\section*{Introduction}
Artificial skyglow is the part of artificial light at night (ALAN) that is scattered or reflected within the atmosphere and directed back towards the Earths surface \cite{Aube:2015}. It is one form of (indirect) light pollution \cite{Riegel1973}. ALAN has grown to a global phenomenon with recent growth rates on the order of by 3-6$\%$ per year \cite{Hoelker:2010_b}, with further increase to be expected \cite{Kyba2014} by efficient solid state lighting technology \cite{Pust2015}. Recent research on ALAN has found that direct and indirect light pollution can negatively affect flora, fauna and human well-being \cite{Stevens:2015}.

Ecological light pollution \cite{book:rich_longcore} (ELP) is the overarching term describing the negative effect of ALAN on the environment. For an introduction, see the review \cite{Longcore2004} and book by Rich and Longcore \cite{book:rich_longcore}. Recent work on ELP include studies on marine turtles \cite{thums2016}, bats \cite{lewanzik2014artificial}, insects \cite{degen2016street}, plants \cite{bennie2016plants, somers2016light} and microorganisms \cite{holker2015microbial}. Furthermore, concerns that ELP can affect whole ecosystems and biodiversity have been raised \cite{Hoelker:2010_a, Gaston:2015_ptb, Kyba:2013_LE}, especially for skyglow as it can cause long range effects \cite{Kyba:2013_LE}. Studies of skyglow in the context of ELP are very sparse, with the exemption of seminal work by Moore on zooplankton \cite{Moore:2000}.

This is mainly owing to the fact that the status-quo of light at night in general and of ALAN (including skyglow) is not well known. Skyglow is highly dynamic with changing atmospheric conditions. Clouds can dramatically increase ALAN in urban areas \cite{Kyba:2011_sqm, Kyba:2012_mssqm, aube2016spectral} but can also reduce ALAN in rural areas \cite{Jechow2016, Ribas2016clouds} as shown by experimental work. Theoretical work is in development but sparse so far \cite{Kocifaj:2014_cloud_theory}. To understand skyglow and ELP in detail, truly interdisciplinary studies are necessary. As a basis for this, it is essential to first fully know the availability of all light at night in the environment \cite{Spitschan2016}. This should include unpolluted sites as well as heavily polluted sites and all states in between, at best at all weather conditions.

Skyglow can be determined from the ground as well as from space. Satellite monitoring \cite{hillger2013first, Katz2016} allows for global coverage with reasonable spatial resolution and revisiting times. In combination with radiative transfer models such data can be used to infer the status of skyglow worldwide, as demonstrated earlier with the world atlas of artificial night sky brightness \cite{Cinzano:2001_wa}, that recently underwent a major update\cite{falchi2016WA}. Other spaceborn sources are (individual) images from astronauts (typically from cities) with improved spatial resolution \cite{kyba2014astroimage}. However, with space-born and airborne methods only the direct upwelling part of ALAN can be measured, while the downwelling part, most relevant for ELP, cannot be accessed directly. Furthermore, atmospheric conditions and especially clouds prevent any satellite data acquisition, rendering it a clear-weather-only method.

Therefore, ground based measurements are still essential to evaluate skyglow and its impact on the environment. Inexpensive single channel devices such as the sky quality meter (SQM, Unihedron, Ontario, Canada) as handheld or permanently installed devices have allowed long-term monitoring with high temporal resolution \cite{Jechow2016, Ribas2016clouds}, the conduction of local surveys \cite{Biggs:2012}, the establishment of local networks \cite{Bara2016} and global comparison \cite{Kyba:2015_isqm}. While the SQM has evolved as a standard device, alternatives like the STARS4ALL night sky brightness photometer have been developed \cite{zamoranostars4all}. However, with both approaches only the luminance (or radiance) at zenith is measured. The zenith luminance is not a good proxy for the total light available at a site, and it is not straightforward to convert it to horizontal or scalar illuminance, commonly used by biologists, as the night sky brightness is usually non-uniform \cite{duriscoe2016photometric, kocifaj2015zenith}. Commercial luxmeters have the drawback that they are either not sensitive enough or, when calibrated for low light levels, are relatively expensive. The IYA Lightmeter could be an affordable alternative but is discontinued and requires self calibration \cite{muller2011measuring}. All single channel devices suffer from the fact, that the spatial distribution and the origin of light pollution cannot be resolved.

Information of the spatial distribution of the night sky brightness is relevant for ELP studies and can be obtained from imaging sensors as nicely outlined by Duriscoe \cite{duriscoe2016photometric}. Several approaches to measure light pollution with imaging sensors exist, like the all-sky transmission monitor (ASTMON) \cite{aceituno2011all, zamorano2016testing}, CCD cameras with sophisticated mechanics to built all-sky mosaics like the US National Park Camera \cite{Duriscoe:2007} as well as other custom built camera systems \cite{rabaza2014new}. However, while these solutions are without question very good options for detailed and precise monitoring of astronomical light pollution on clear nights,  they are rather complicated in handling (e.g. by using multiple filters or requiring photometric calibration with celestial objects), relatively slow in data acquiring (e.g. about 40 minutes for an all-sky mosaic \cite{Duriscoe:2007}) or require extensive data processing and or calibration \cite{rabaza2014new}. 

The importance of imaging devices for light pollution measurements has been pointed out earlier \cite{duriscoe2016photometric, rabaza2010all, solano2016allsky}. However, it is not widely recognized in the community investigating ELP, with the exception of work on marine turtle hatchlings \cite{pendoley2012}. We assume that this is mainly owing to the fact that the threshold of using the aforementioned systems by biologists in the field is high due to their complexity. Recently, commercial digital single lens reflex (DSLR) cameras with fisheye lenses have been used for night sky studies \cite{Kollath:2010, Kocifaj:2015, solano2016allsky, Jechow2016, Jechow2017}. These cameras come with reasonable factory calibration and promise to allow data acquisition by non specialists as they do not require special training or complex knowledge about photometry or electronics. The huge public demand for imaging devices in consumer electronics like smartphones, has also resulted in dropped prizes for stand alone cameras, making professional level DSLR cameras cheaper than calibrated luxmeters.

Here we show that off-the-shelf commercial DSLR cameras are well suited for measuring the night sky brightness in the field under overcast as well as clear conditions. The simple and fast data acquisition allowed us to perform two rapid transects from the town of Balaguer towards Montsec Astronomical Park in Spain, covering more than 20 km in about 2 hours. We were able to map the reach of skyglow from the peri-urban area into the remote area. From the data, we can obtain illustrative luminance maps, extract the zenith luminance as well as horizontal and scalar illuminance (see methods), giving a comprehensive set of data that cannot be inferred easily from single channel sensors or satellite imagery, especially not under cloudy conditions. We find that with the presence of clouds, areas that reach near-natural zenith brightnesses under clear sky conditions experience the same light levels as peri-urban area under clear conditions. Such a shift of anthropogenic skyglow into pristine areas has not been investigated at this level of detail with imaging sensors. We judge that the method is widely applicable for field studies up to the global level as it relies on off-the-shelf equipment, making it an ideal choice for non-specialists in photometry. 

\section*{Results: Transect near Montsec Astronomical Park and the Spanish town of Balaguer}
\begin{figure}[h]
\centering
\includegraphics[width=\columnwidth]{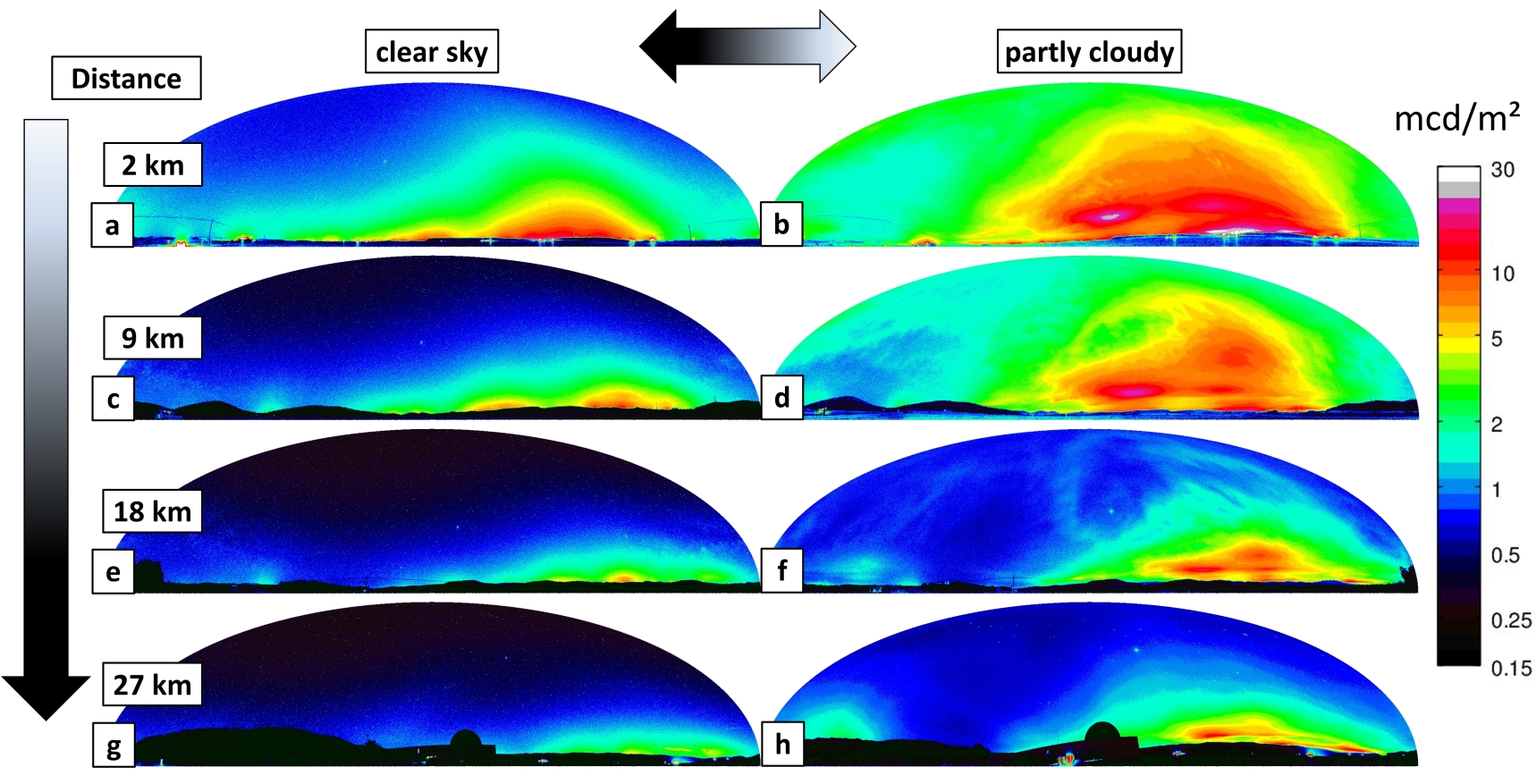}
\caption{Luminance maps of the night sky in Hammer–Aitoff equal area projection calculated from all-sky images obtained along the transects from the town of Balaguer towards Montsec Astronomical Park in May 2016. The left column (a, c, e, g) shows data obtained on the night of May 3rd/May 4th with a clear sky and the right column (b, d, f, h) shows data obtained at the night of May 5th/May 6th with partly cloudy sky. The upper row (a, b) was acquired at 2 km distance, the second row (c, d) at 8.5 km distance, the third row (e, f) at 17.7 km distance to the city center of Balaguer. The lower row was obtained at the Astronomical Park of Montsec at 27 km distance to Balaguer. The full data set is shown in the Appendix.}
\label{cloud_comp}
\end{figure}

Figure \ref{cloud_comp} shows 8 (of 18 total) luminance maps of the night sky obtained during two transects from the town of Balaguer towards Montsec Astronomical Park in May 2016. The images are shown in Hammer–Aitoff equal area projection and were calculated from all-sky images obtained with a commercial DSLR camera. In the left column (a, c, e, g) of Fig. \ref{cloud_comp} data from a night with clear sky (May 3-4) is shown. In the right column (b, d, f, h) of Fig. \ref{cloud_comp} data from a night with partly cloudy sky (May 5-6) is shown. The upper row (a, b) shows data taken at 2 km distance, the second row (c, d) data taken at 8.5 km distance and the third row (e, f) data taken at 17.7 km distance to the town center of Balaguer. The lower row shows data taken at Astronomical Park of Montsec at 27 km distance to Balaguer. The full data set, consisting of 18 all-sky luminance maps, is shown in the Appendix in Fig. \ref{app1} and Fig. \ref{app2}. The measurement locations, topography and other artificial skyglow data from the region are described in the methods section.

From the clear sky luminance map obtained at 2 km distance shown in Fig. \ref{cloud_comp} a), the light dome originating from the skyglow of the town Balaguer is clearly visible as a big hump on the right. The skyglow from the city of Lleida (24.8 km distance to this position) is apparent by a smaller hump in the center of the image (left of the Balaguer). Within the clear sky data set, these two sky domes diminish with progressing distance and merge into one broader hump confined to the horizon.

From Fig. \ref{cloud_comp}, the impact of the clouds on the luminance of the whole sky becomes immediately apparent. For all distances, the partly cloudy night (right column) has a higher luminance than the corresponding clear night (left column). Furthermore, the impact of the sky dome from Lleida seems to have a higher impact for cloudy than for clear nights, as can be seen by a spot-like feature in the center of Fig. \ref{cloud_comp} b) and d). Furthermore, some of the skyglow that is masked by e.g. mountains on clear nights (as it is confined to the horizon) becomes apparent for partly cloudy conditions. This can be seen when examining the lowest row in Fig. \ref{cloud_comp} and comparing the luminance distribution on the left side of the images.  Also, without any further data analysis, Fig. \ref{cloud_comp} a) appears most similar to Fig. \ref{cloud_comp} f), with the latter appearing slightly darker.

\begin{figure}[h!]
\centering
a)\includegraphics[width=80mm]{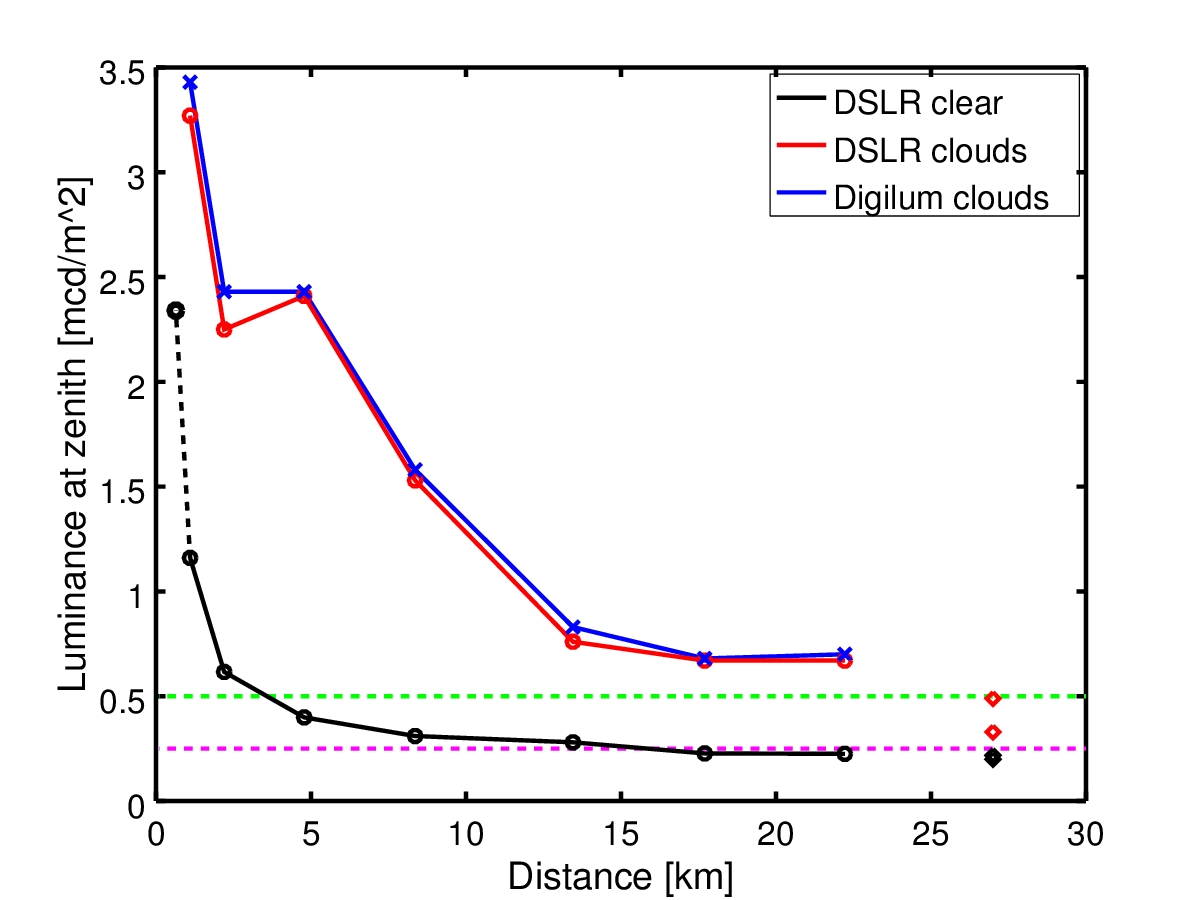}
b)\includegraphics[width=80mm]{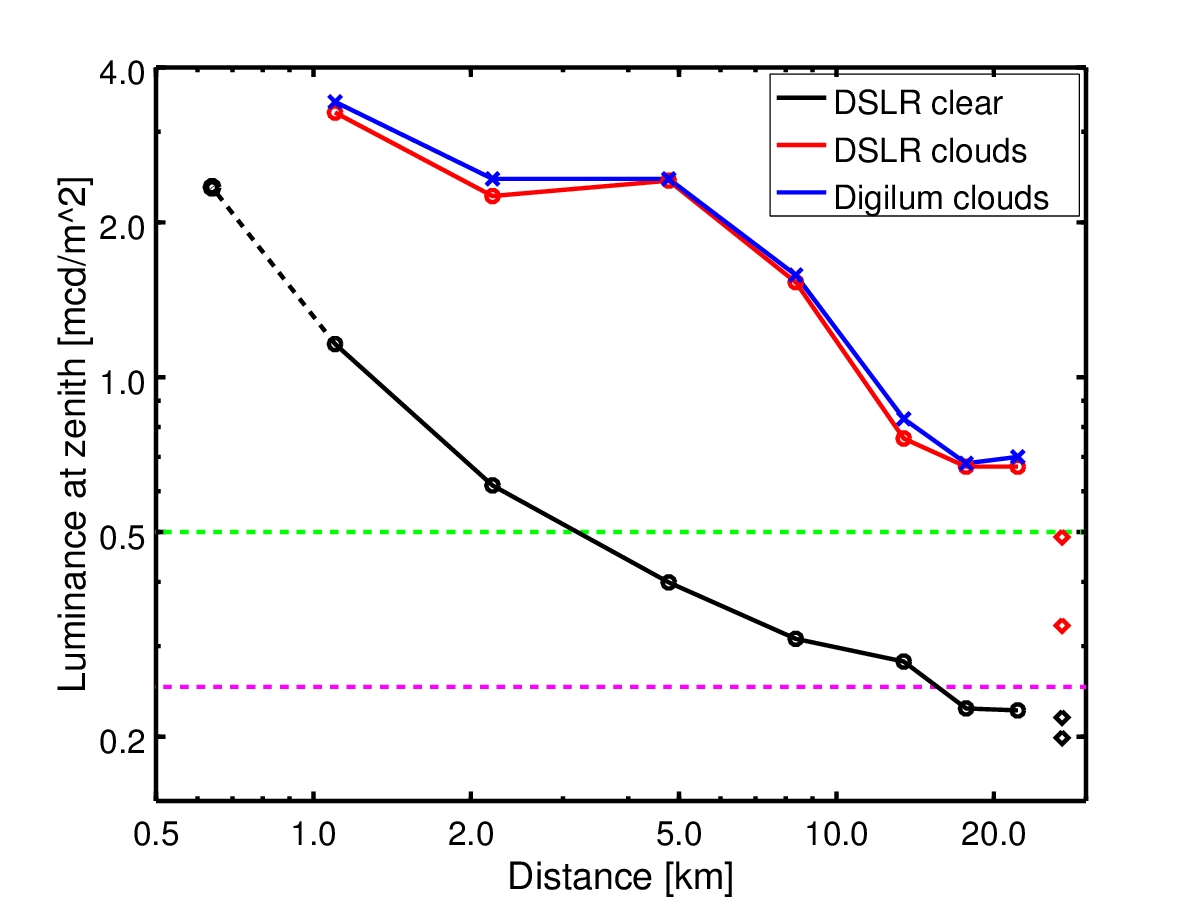}
\caption{Luminance at the zenith as a function of the distance from the city center of Balaguer measured at two different nights taken at the transects with clear sky (black open circle and solid line) and partly cloudy sky (red open circles and solid line) in a linear scale a) and a double logarithmic scale b). Additionally data from the Astronomical park at 27.2 km distance is shown as black (clear sky) and red (partly cloudy sky) diamonds in both plots, respectively. For the clear sky case, an extra data point closer to the town center (600m) was taken, which is indicated by the black dashed line. For the partly cloudy night, comparative data obtained with the Digilum luminance meter is shown in blue. The dashed magenta line indicates the value of a ``typical'' clear sky (0.250 mcd/m$^2$) and the green dashed line a value twice as much (0.50 mcd/m$^2$).}
\label{Bal_dist}
\end{figure}

From the all-sky images, the zenith luminance, $ L_{v, zen}$, was calculated (see methods). In Fig. \ref{Bal_dist}, $ L_{v, zen}$ is plotted as a function of the distance from the town of Balaguer  on a linear a) and on a double logarithmic scale b). The black open circle and solid line show the transect data for the clear night and the red open circles and solid line show the data from the transect for the partly cloudy night.  The diamonds in both plots show data taken at the Astronomical park site at 27 km distance to Balaguer (lower row in \ref{cloud_comp}) with black diamonds for clear sky and red diamonds for partly cloudy sky. For the clear sky an extra data point closer to the town center was taken, which is indicated by the black dashed line. Additionally, the zenith luminance values measured with the Digilum luminance meter (Optronik GmbH, Berlin, Germany) are shown for the partly cloudy night (blue crosses and solid line). The individual values from the whole data set are summarized in table \ref{table_zenith}. To guide the eye, the dashed magenta line indicates the value often referred to be the luminance of a ``typical'' clear sky (0.250 mcd/m$^2$) \cite{falchi2016WA}, please note that there are other values in use \cite{duriscoe2016photometric}) and the green dashed line indicates a value twice as much (0.50 mcd/m$^2$).

For the clear sky case, the $ L_{v, zen}$ values drop from 1.16 $\pm$ 0.12 mcd/m$^2$ at 1 km distance to the center of Balaguer to values as low as 0.20 $\pm$ 0.02 mcd/m$^2$ for the Astronomical Park at 27 km distance. The value of 0.62 $\pm$ 0.06 measured at 2 km is already in the vicinity of 0.50 mcd/m$^2$ (green line), while the value at 8.4 km is only 25$\%$ above the 0.25 mcd/m$^2$ reference (magenta line).

With the presence of clouds, $L_{v, zen}$ values increase up to a factor of 6 at the individual sites (see table \ref{table_zenith} right hand column), with $L_{v, zen}$  dropping from 3.27 $\pm$ 0.33 mcd/m$^2$ at 1 km distance to 0.49 $\pm$ 0.05 mcd/m$^2$ at the Astronomical Park. Later in the night, with the sky clearing up, $L_{v, zen}$ reaches 0.33 $\pm$ 0.04 mcd/m$^2$ at the latter, most distant site. It is interesting to note, that all values obtained along the transect (stop 1-7) are higher than that obtained at 2 km distance for the clear night. Furthermore, the $L_{v, zen}$ value at 5 km distance is as high as measured inside of the town (600m to center) for the clear night.

\begin{table}[h!]
  \centering
  \caption{List of the zenith luminance values at the different locations and the two nights (see methods for details). Please note that the values for the last measurement site show data from earlier in the night (8$^{*}$) and later in the night (8$^{**}$).}
  \label{table_zenith}
  \begin{tabular}{cccccc}
    \toprule
   Stop Nr. & Distance &  &Zenith Luminance $ L_{v,zen}$ [mcd/m$^2$]& &Ratio   \\
 &  & DSLR clear & DSLR clouds  & Digilum clouds& $L_{v,zen,cloud}/ L_{v,zen,clear}$\\
    \toprule
1 & 1.1 km & 1.16 $\pm$ 0.12& 3.27 $\pm$ 0.33 & 3.43 $\pm$ 0.34 & 2.8 \\
2 & 2.2 km & 0.62 $\pm$ 0.06 & 2.25 $\pm$ 0.23 & 2.43 $\pm$ 0.24 & 3.7\\
3 & 4.8 km & 0.40 $\pm$ 0.04& 2.41 $\pm$ 0.24 & 2.43 $\pm$ 0.24 & 6.0\\
4 & 8.4 km & 0.31 $\pm$ 0.03& 1.53 $\pm$ 0.15 & 1.58 $\pm$ 0.16 & 4.9\\
5 & 13.5 km& 0.28 $\pm$ 0.03& 0.76 $\pm$ 0.08 & 0.83 $\pm$ 0.08 & 2.7\\
6 & 17,7 km& 0.23 $\pm$ 0.02& 0.67 $\pm$ 0.07 & 0.68 $\pm$ 0.07 & 2.9\\
7 & 22.2 km& 0.22 $\pm$ 0.02& 0.67 $\pm$ 0.07 & 0.70 $\pm$ 0.07 & 3.0\\
8$^{*}$ &27.2 km& 0.22 $\pm$ 0.02& 0.49 $\pm$ 0.05  & -- & 2.2\\
8$^{**}$ &27.2 km& 0.20 $\pm$ 0.02&  0.33 $\pm$ 0.04 & -- & 1.7\\
    \toprule
  \end{tabular}
\end{table}

\begin{figure}[h!]
\centering
a)\includegraphics[width=80mm]{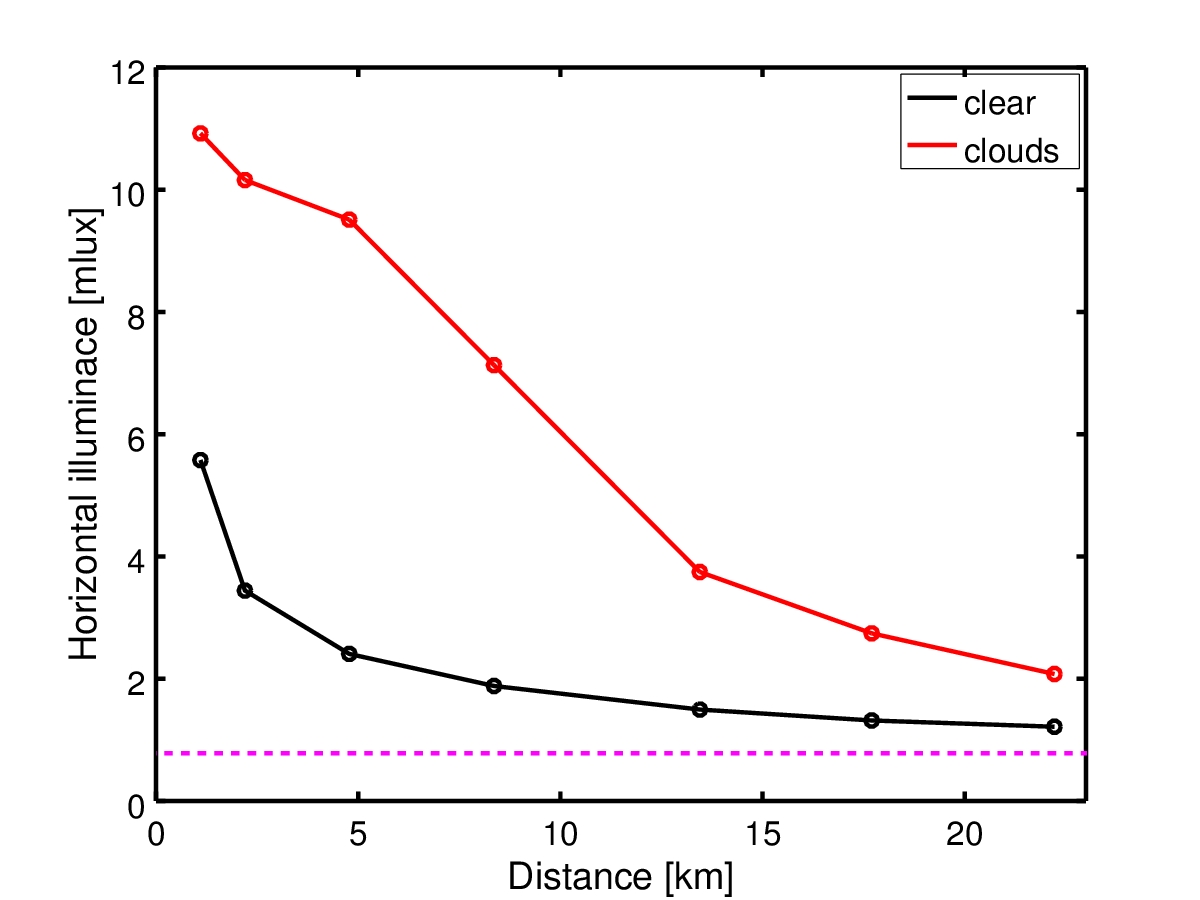}
b)\includegraphics[width=80mm]{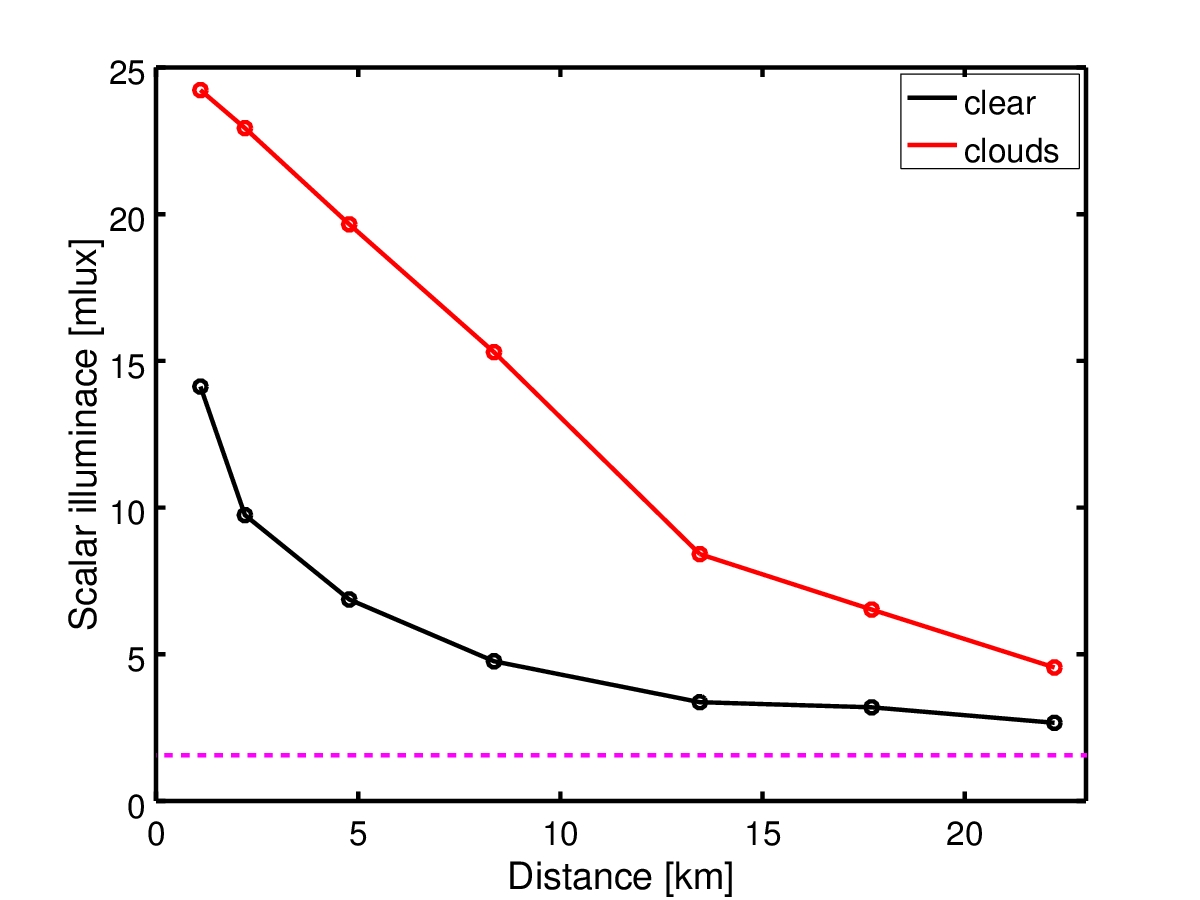}
\caption{a) Horizontal and b) scalar (hemispheric) illuminance as a function of the distance from the city center of Balaguer measured for two different nights with clear sky (black) and partly cloudy sky (red). Magenta lines show estimated ideal values (see methods).}
\label{Bal_ill}
\end{figure}

\pagebreak
It is further possible to calculate horizontal illuminance, $E_{v,hor}$, and scalar illuminance, $E_{v,scal,hem}$, (here for the hemisphere) in the commonly used unit $lux$ from the all-sky data, as the zenith luminance might not be representative for all light incident at a specific site (see methods for detail). Figure \ref{Bal_ill} a) shows the horizontal illuminance for clear conditions (black) and partly cloudy conditions (red), with a hypothetical ideal natural-sky illuminance value of $E_{v,hor,ideal} \approx 0.78 mlux$ \cite{falchi2016WA} (again other values are reported in literature \cite{kocifaj2015zenith}, see methods and discussion) indicated by the horizontal dashed magenta line. The trend is the similar as for the zenith luminance but weaker. For example, the ratios between cloudy and clear nights are all lower than for the zenith luminance. Here, the value of  $E_{v,hor}$ at 2.2 km distance for the clear night is in the same range as  $E_{v,hor}$ at 13.5 km for the cloudy night.

Figure \ref{Bal_ill} b) shows the scalar illuminance as a function of distance from the center of Balaguer, again  for clear conditions (black) and partly cloudy conditions (red) as well as with the ideal scalar illuminance, $E_{v,scal,hem,ideal} \approx 1.56 mlux$, indicated by the horizontal dashed magenta line. The overall trend is about the same as for the horizontal illuminance.

\begin{table}[h!]
  \centering
  \caption{List of the illuminance values for the two nights at the different locations along the transect (see methods for details). Please note that the values for the last measurement site show data from earlier in the night (8$^{*}$) and later in the night (8$^{**}$).}
  \label{table_ill}
  \begin{tabular}{cccccccc}
    \toprule
   Stop Nr. & Distance & Hor. Illuminance & [mlux]& &  Scal. Illuminance&[mlux] \\
 &  & clear & clouds  & Ratio & clear & clouds & Ratio\\
    \toprule
1 & 1.1 km & 5.6 $\pm$ 0.6& 10.9 $\pm$ 1.1 & 1.9& 14.1 $\pm$ 1.4& 24.2 $\pm$ 2.4 & 1.7 \\
2 & 2.2 km & 3.4 $\pm$ 0.3& 10.2 $\pm$ 1.0 & 3.0& 9.7 $\pm$ 1.0& 22.9 $\pm$ 2.3 & 2.4\\
3 & 4.8 km & 2.4 $\pm$ 0.2& 9.5 $\pm$ 1.0 & 4.0& 6.9 $\pm$ 0.7& 19.6 $\pm$ 2.0 & 2.8\\
4 & 8.4 km & 1.9 $\pm$ 0.2& 7.1 $\pm$ 0.7 & 3.7& 4.8 $\pm$ 0.5& 15.3 $\pm$ 1.5 & 3.2\\
5 & 13.5 km& 1.5 $\pm$ 0.2& 3.7 $\pm$ 0.4 & 2.5& 3.4 $\pm$ 0.3& 8.4 $\pm$ 0.8 & 2.5\\
6 & 17,7 km& 1.3 $\pm$ 0.1& 2.7 $\pm$ 0.3 & 2.1 & 3.2 $\pm$ 0.3& 6.5 $\pm$ 0.6 & 2.0\\
7 & 22.2 km& 1.2 $\pm$ 0.1& 2.1 $\pm$ 0.2 & 1.7 & 2.7 $\pm$ 0.3& 4.6 $\pm$ 0.5 & 1.7\\
8$^{*}$ &27.2 km& 1.2 $\pm$ 0.1 & 2.8 $\pm$ 0.3  & 2.3 & 2.3 $\pm$ 0.2& 5.0$\pm$ 0.5 & 2.2\\
8$^{**}$ &27.2 km& 1.1 $\pm$ 0.1 & 2.5 $\pm$ 0.3 & 2.3 & 1.6 $\pm$ 0.2& 3.5 $\pm$ 0.4 & 2.2\\
    \toprule
  \end{tabular}
\end{table}

\pagebreak
%%%%%%%%%%%%%%%%%%%%%%%%%%%%%%%%%%%%%%%%%%
\section*{Discussion}
One of the big advantages of this method is that it makes it possible to compare different physical quantities acquired simultaneously in a single image. The photometric data from the all-sky images allowed us to extract zenith luminance, horizontal illuminance and (hemispheric) scalar illuminance simultaneously. Along the transects, we could investigate the impact of clouds on the amplification of skyglow at individual sites, and map the increase in the reach of skyglow. The most dramatic change in individual site brightening and extension of the reach of skyglow was observed for the zenith luminance levels, while for the illuminance measurements this shift was lowered. Zenith luminance observed on clear nights can underestimate the degree to which a site is exposed to skyglow under all weather conditions. While the zenith appears near-natural, there can be significant skyglow at the horizon. With the presence of clouds, skyglow can ``creep'' into the field of view of the measurement device.

This is apparent for example for the brightness values obtained at 8.4 km distance to Balaguer in \ref{cloud_comp} c) and d). At this site, the zenith luminance for the clear night $L_{v, zen}$=0.31 $\pm$ 0.03 mcd/m$^2$ is only 24$\%$ above the value of 0.250 mcd/m$^2$, appearing as almost natural sky. The horizontal illuminance at the clear night $E_{v,hor}$=1.9 $\pm$ 0.2 mlux is already 2.4 times as much as the reference value of $0.78 mlux$, while the (hemispherical) scalar illuminance $E_{v,scal,hem}$ is more 3.1 times the dark sky value of $1.56 mlux$. The amplification factor for clouds is 4.9 for the zenith luminance, resulting in a $L_{v, zen}$ value 6.1 times brighter compared to the natural sky value of 250 mcd/m$^2$. For the horizontal illuminance, the amplification factor is 3.7, resulting in an 8.9 times higher value of $E_{v,hor}$ with respect to the reference case. For $E_{v,scal,hem}$ the amplification factor is 3.2 resulting in 9.9 higher value compared to the ideal clear sky case.

In the context of ELP, there are hardly any animals that just sense the zenith brightness alone. Thus illuminances are a better proxy for ELP. Therefore, the amount of light available for ecological effects is underestimated by a factor of 2 and 2.5 by SQMs (or similar devices) for the clear sky case, when inferring illuminance levels from zenith brightness compared to measured illuminances. On the other hand, the brightening effects of clouds can be overestimated by single channel devices.

For all three photometric quantities, the reach of skyglow was enhanced with the presence of clouds. An extreme shift was observed in particular for the zenith luminance $L_{v, zen}$, with higher luminance observed at 22.2 km from the town for the partly cloudy night than were seen at 2.2 km for the clear night. For the horizontal illuminance, this ``distance-effect'' is slightly reduced. For the partly cloudy night, $E_{v,hor}$ was higher at 13.5 km than was observed at 2.2 km for the clear night. For $E_{v,scal,hem}$, the size of the difference was further reduced.

The other big advantage of the method is its potential wide applicability. So far, all-sky imaging tools \cite{duriscoe2016photometric, rabaza2010all, solano2016allsky, Kollath:2010, Kocifaj:2015, solano2016allsky, Jechow2016, Jechow2017} have been almost exclusively used by astronomers, physicists, and engineers, with the notable exception of scientists working with marine turtles \cite{pendoley2012}. However, spatial information might be vital for understanding the effects of ELP, as some species possibly use spatial features for orientation \cite{dacke2013dung}. We assume that wide application of imaging tools was so far hampered by the complexity of existing methods. Operation of the commercial DSLR camera is straightforward, and well suited for fast acquisition during fieldwork with a minimum of training necessary. The camera is mobile, yet rigid and reliable, and we have so far never experienced any equipment failure. In our case, reaching a vantage point in the hilly terrain from the road sometimes took longer than the data acquisition itself. The data analysis is not too complicated, and the software as well as the code used here is freely available. However, one drawback is that the light level values measured are not currently immediately available as they would be with some other instruments.

While this work focuses on spatially resolved luminance measurement of the whole hemisphere, the method can be extended to use the three color channels of the camera to obtain further spectral information\cite{solano2016allsky}. Recent studies have determined the spectral irradiance with spatially single channel sensors of the sky \cite{Spitschan2016} and even under water \cite{Tamir2017} showing the dramatic alteration of the night light climate. Other recent work includes the use of hyperspectral imaging systems to determine the (direct) light available in an urban setting \cite{Alamus2017}.

%%%%%%%%%%%%%%%%%%%%%%%%%%%%%%%%%%%%%%%%%%
\section*{Conclusions}

We have shown that DSLR cameras with fisheye lenses are well suited for all-sky field studies in the context of ELP under both clear and cloudy conditions. By performing two transects from a peri-urban area into a remote area, we observed a shift in the reach of skyglow under clear and partly cloudy conditions. The data support the hypothesis that clouds can dramatically extend the reach of skyglow, resulting in ALAN penetrating areas that appear pristine on clear nights, and filling the gaps between illuminated areas. At our study site, zenith luminance levels at 22 km from a small city on a cloudy night are similar to those at 2 km on clear night. This trend is also observed in the more ELP-relevant illuminance levels, with slightly lower amplification and extension. We further find evidence that in some scenarios, the zenith brightness is not a good proxy for the light available at a site, supporting concerns raised recently \cite{kocifaj2015zenith}.

The strength of all-sky imaging is that several photometric quantities can be extracted from one image, and that individual parts of the night sky can be investigated. This might be essential for studies on animal behavior in the context of ELP\cite{dacke2013dung}. This method uses off-the-shelf equipment, and could be easily adopted by the ELP community. This would hopefully result in increased and global data on the state of skyglow under cloudy as well as clear conditions. Single site studies would be useful, as well as other transect data. The method may also lend itself to long-term monitoring, and possibly data acquisition by citizen scientists\cite{Kyba:2013_GaN}.

%%%%%%%%%%%%%%%%%%%%%%%%%%%%%%%%%%%%%%%%%%
\section*{Methods}
\subsection*{Study site}
\subsubsection*{Transect from Balaguer to Montsec Astronomical Park, and the status of artificial skyglow in the area}
During the 2016 Stars4All/LoNNe intercomparison campaign held at Montsec Astronomical Park (Parc Astronòmic Montsec, Centre d' Observació de l' Univers: PAM-COU), two transects were conducted on a clear night (May 3rd to May 4th) and a partly cloudy night (May 5th to May 6th). Details on the intercomparison campaign, including other activities and other equipment used, can be found in the campaign report \cite{RibasLonne2017}.

\begin{figure}[h!]
\centering
\includegraphics[width=0.9\columnwidth]{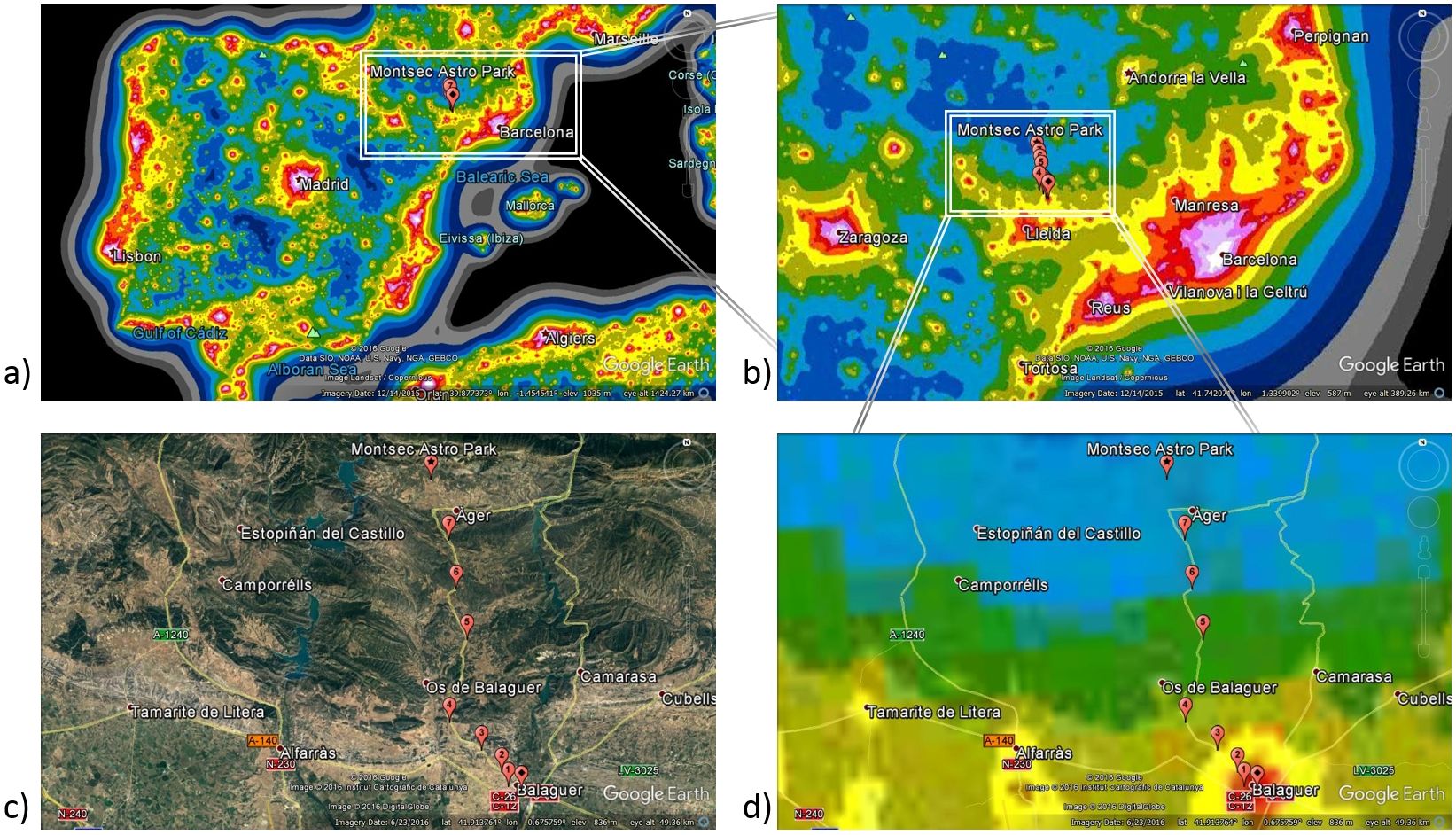}
\caption{Location of the transect near the Spanish town of Balaguer towards Montsec Astronomical Park: Night sky brightness maps (taken from \cite{falchi2016WA}) of a) Spain, b) Catalonia and d) the local area. c) shows a map of the local area.}
\label{WA_map}
\end{figure}

Figure \ref{WA_map} shows the ''World Atlas'' predictions for (clear sky) artificial skyglow \cite{falchi2016WA} for a) the Iberian Peninsula b) Catalonia and surroundings, and d) the area of the transect. In Fig. \ref{WA_map} c) a local map is shown. The transects were started at the outskirts of Balaguer (41.791100N, 0.797494E, 273m elevation) and were conducted along the highway C-12 leading North towards Àger and PAM-COU, respectively. The last stop on the route was at Port d' Àger: (41.979070N, 0.750630E, 908m elevation). In between these two points, measurements at five further stops have been made for each of the two nights. The seven stops are labeled with numbers in the map. Additional indicators in Fig. \ref{WA_map} c) and d) show the center of Balaguer and the location of the PAM-COU, where additional measurements were made synchonously with the transect. Details on all measurement sites can be found in table \ref{table_sites}.

\begin{table}[h!]
  \centering
  \caption{List of the locations along the transect from Balaguer to Port d' Àger. Stationary observations were performed during the entire period of the transect from the final location (Parc Astronòmic Montsec, PAM-COU).}
  \label{table_sites}
  \begin{tabular}{ccccc}
    \toprule
   Stop Nr. & Name & Position & Elevation & Distance to Balaguer\\
    \toprule
1 & Balaguer outskirts & 41.7911N, 0.7975E & 273m & 1.1 km\\
2 & Cemetary & 41.8024N, 0.7916E & 287m & 2.2 km\\
3 & Hill-1 & 41.8120N, 0.7725E & 395m & 4.8 km\\
4 & Plain & 41.8426N, 0.7418E & 412m & 8.4 km\\
5 & Village & 41.9038N, 0.7640E & 595m & 13.5 km\\
6 & Hill-2 & 41.9417N, 0.7560E & 662m &17,7 km\\
7 & Port d' Àger & 41.9791N, 0.7506E & 908m & 22.2 km\\
8$^{*}$ & Parc Astronòmic Montsec & 42.0248N, 0.7368E & 813m & 27.2 km\\
    \toprule
  \end{tabular}
\end{table}

The topography of the area is dominated by the mountains of the Montsec region. Along the transect, the highway C-12 climbs from 273m elevation (stop 1) to 908m (stop 7). The dominant sources for artificial skyglow are the light domes of Balaguer (27.2 km from PAM-COU) and Lleida (ca. 46 km from PAM-COU), while the larger light dome of Barcelona (ca. 140 km from PAM-COU) is also apparent near the horizon. According to Falchi et al. \cite{falchi2016WA} the calculated clear sky lumnance in the center of Balaguer is 1.07-1.96 mcd/m$^2$ and falls off towards PAM to 0.2-0.23 mcd/m$^2$ (see Fig. \ref{WA_map}). However, we want to point out that this global model does not include local characteristics such as the topography or vegetation cover of the landscape, yet.

Fortunately for astronomical observations, the PAM-COU (measurement location 8) is located in a valley surrounded by hills and mountains, usually shielding the artificial skyglow present at the horizon. For the majority of clear nights, the zenith night sky brightness is near-natural with values of 22 mag$_{SQM}$/arcsec$^2$. On cloudy nights the zenith night sky brightness can either decrease or increase, as shown by an in depth-study linking SQM values and cloud height at PAM-COU \cite{Ribas2016clouds}.
\begin{figure}[h!]
\centering
\includegraphics[width=85mm]{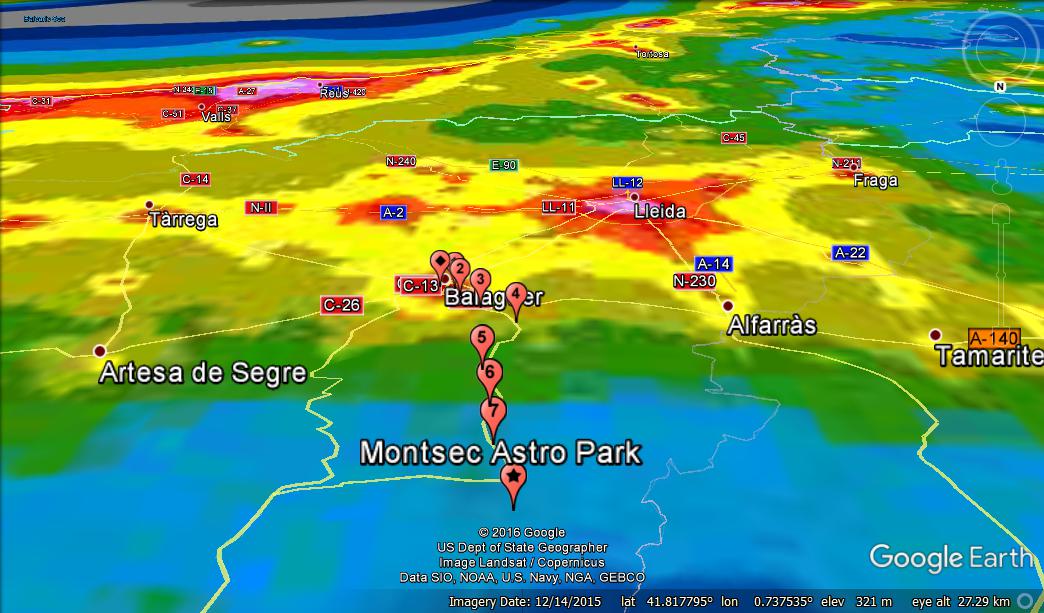}
\caption{Night sky brightness map (taken from \cite{falchi2016WA}) as seen from Parc Astronòmic Montsec towards Balaguer.}
\label{PAM_map}
\end{figure}

\subsubsection*{Weather conditions}
On 3 May 2017, the sky was clear during the day with some cirrus, still significant around sunset but disappearing by around nautical twilight. No clouds were picked up by ceilometer at PAM after 10:20 pm local time (8:20 pm UTC). There were no clouds at all by astronomical twilight, and this condition lasted through the entire night, including during the morning twilight. At the end of the night, the humidity level increased to 78$\%$ (PAM-COU Weather Station).

On May 5th, the sky became progressively cloudier throughout the day, with high clouds. The entire evening was characterized by high thin clouds. At PAM, stars were visible, and there was a moisture layer up to the top of the mountain which allowed scattered light from very remote cities to be seen. The scattering from upward-directed lights from the villages to the North of the mountain was conspicuous. The sky was covered the whole night with medium and high clouds. The ceilometer measured cloud heights around 6 km for the relevant measurement period. Early in the morning low clouds also appeared, according to ceilometer measurements.

\subsection*{DSLR camera}
The all-sky images were obtained with a commercial DSLR camera (Canon EOS 6D). This camera has a full-frame CMOS sensor with 20.2 Megapixel (5496 x 3670 pixels) and allows ISO settings from 100 to 25,600 and shutter speed ranges of 1/4,000 to 30 s without extra equipment. It has a built in GPS sensor. The camera was operated with a circular fisheye lens (Sigma EX DG with 8 mm focal length). The lens was focused to a bright celestial object in the live view of the camera. If celestial objects are masked by clouds, focusing can be done at an artificial light source at large distance (e.g. on the horizon). 

To be both mobile in the field yet acquire still images, the camera was mounted on a tripod. To acquire all-sky images, the camera was aligned with the center of the lens oriented towards the zenith. For the light levels present during the two nights, ISO 1600 was used for images closer to the town and ISO 3200 for images further away from the town. The shutter speed was varied between 15 s and 30 s along the transect, while a long exposure of 120 s was used at the Montsec Astronomical Park location. For the cloudy night, several images were obtained and averaged.

The method applied here of using commercial DSLR cameras with fisheye lenses is simple as well as robust, although not as rigid as a commercial outdoor or even waterproof light measuring devices like luxmeters. The point and click ability offers great opportunities, especially if measurements have to be acquired quickly. Furthermore, commercial DSLR cameras are now quite affordable (in the same price range as a calibrated luxmeter or PAR sensor). The spatially resolved data allow for the extraction of several parameters acquired at the same time (see below).

\subsection*{DiCaLUM software}
Image processing was performed using DiCaLum (Version 1.01, by Zolt{\'a}n Koll{\'a}th) (see ZK IJSL). The code is programmed in GNU Octave and relies on the free software DCRaw. The RAW images are read out, and the radiance or luminance can be obtained from the pixel values of the three different channels. In the present version, the software uses the green channel, which has significant overlap with the V($\lambda$) curve referenced to human photopic vision (cite A. Hänel). The pixel values from the green channel are then used to map the luminance values for each individual pixel.

Although highly standardized, DSLR cameras and lenses still require thorough calibration, especially for the vignetting of the fisheye lens. DiCaLum has several cameras and fisheye lenses in its library. Initial calibration was done in the laboratory, while some cameras were cross calibrated in several intercomparison campaigns \cite{KybaLonne2015, RibasLonne2017}. For further reading on camera calibration we refer to \cite{Kocifaj:2015, solano2016allsky} where the method is described in more detail.

A common measure to quantify how much light is available in the environment is horizontal illuminance, the total luminous flux incident on a flat horizontal surface (here the Earth's surface), per unit area. When assuming the sky to be a hemisphere, the horizontal illuminance is defined as:
\begin{equation}
E_{v,hor}=\int_{0}^{\pi \over 2}\int_{0}^{2\pi} L_{v,sky}(\theta, \phi) sin\theta cos\theta d\theta d\phi,
\label{eq_hor}
\end{equation}
with $\theta$ being the zenith angle, $\phi$ being the azimuth angle, and $L_{v,sky}(\theta, \phi)$, being the luminance of the sky. The SI unit is $lux$. The horizontal illuminance is sometimes also referred to be cosine corrected. Commonly, the horizontal illuminance is inferred from the zenith luminance, $L_{v,sky,zen}$. When assuming a uniform sky luminance $L_{v,sky}(\theta, \phi) = const.=L_{v,zen}$, we can treat $L_{v,zen}$ as a constant and do not have to integrate over it, while the integral in equation \ref{eq_hor} simply becomes $2\pi$ from integration over the azimuth and 0.5 from integration over the zenith angle, giving
\begin{equation}
E_{v,hor} \approx \pi \cdot  L_{v,sky,zen}
\label{eq_zenill}.
\end{equation}
With the commonly used assumption of a typical clear night sky zenith luminance of $L_{v,zen,ideal}$ = 0.25 mcd/m$^2$, we can approximately infer a typical illuminance for this case: $E_{v,hor,ideal} \approx \pi \cdot  0.25 \mathrm{mcd/m}^2 \approx 0.78$ mlux. As discussed in more detail in \cite{kocifaj2015zenith}, this might not be a valid approximation for most real scenarios, as the sky brightness is rather nonuniform (evident in our own data, see Fig. \ref{cloud_comp}).

The horizontal illuminance may not be the best measure for all organisms affected by ecological light pollution, as for some species the angle of incidence may not be of primary importance. For this case, the total luminous flux incident on a small spherical surface gives the so-called scalar illuminance. For the sky and the all-sky images, it makes sense to define a hemispherical scalar illuminance \cite{duriscoe2016photometric}:
\begin{equation}
E_{v,scal,hem}=\int_{0}^{\pi \over 2}\int_{0}^{2\pi} L_{v,sky}(\theta, \phi) sin\theta d\theta d\phi,
\end{equation}
which does not incorporate the cosine correction.
The DiCaLum code can produce luminance maps, extract zenith luminance, and horizontal and (hemispherical) scalar illuminance, by summing the solid angle weighted luminance values with or without additional cosine correction.

\subsection*{Digilum luminance meter}
The Digilum (Optronik GmbH, Germany) is a custom luminance meter specifically tailored and calibrated for low light level detection at the zenith. It has a 5$^{\circ}$ aperture and a spectral response that is well matched to photopic vision, V($\lambda$). The Digilum was mounted on the back of the car for the second transect during the partly cloudy sky and is constructed on a gimbal, guaranteeing straight pointing to the zenith. Averaged luminance values were obtained for the measurement duration of the corresponding DSLR imaging. A detailed description of dark current subtraction is given in the Stars4All intercomparison campaign report \cite{RibasLonne2017}. The device has been used in other intercomparison measurements before \cite{den2015stability}.

%%%%%%%%%%%%%%%%%%%%%%%%%%%%%%%%%%%%%%%%%%
\section*{Acknowledgments}
We thank all the participants at the Stars4All/LoNNe intercomparison that supported this work, specifically Guillem Marti and Pol Massana for driving the car on the first day of the transect, Constantinos Bouroussis for joining the transect on the second night.

Andreas Jechow is supported by the ILES project funded by the Leibniz Association, Germany (SAW-2015-IGB-1). We thank the Stars4All awareness platform that funded the intercomparison campaign. This article is based upon work from COST Action ES1204 LoNNe, supported by COST (European Cooperation in Science and Technology).
%%%%%%%%%%%%%%%%%%%%%%%%%%%%%%%%%%%%%%%%%%
\section*{Author contributions statement}
AJ conceived the study with input from CCMK and FH. AJ acquired the transect data, analyzed the data and wrote the first draft of the manuscript. ZK acquired the data at Montsec and programmed DiCaLum. HS acquired and analyzied the Digilum measurements. SJR organized and planned the intercomparison campaign in coordination with members of LoNNe WG 1 (including CCMK, ZK, AJ, and others), and SJR also selected the first and the last stop of the transect. The final manuscript was written with input from all authors.

\section*{Competing financial interests}
The authors declare no competing financial interests.

%%%%%%%%%%%%%%%%%%%%%%%%%%%%%%%%%%%%%%%%%%
\newcommand{\beginsupplement}{%
        \setcounter{table}{0}
        \renewcommand{\thetable}{S\arabic{table}}%
        \setcounter{figure}{0}
        \renewcommand{\thefigure}{S\arabic{figure}}%
     }

\beginsupplement
\begin{figure}[h]
\centering
a) \includegraphics[width=50mm]{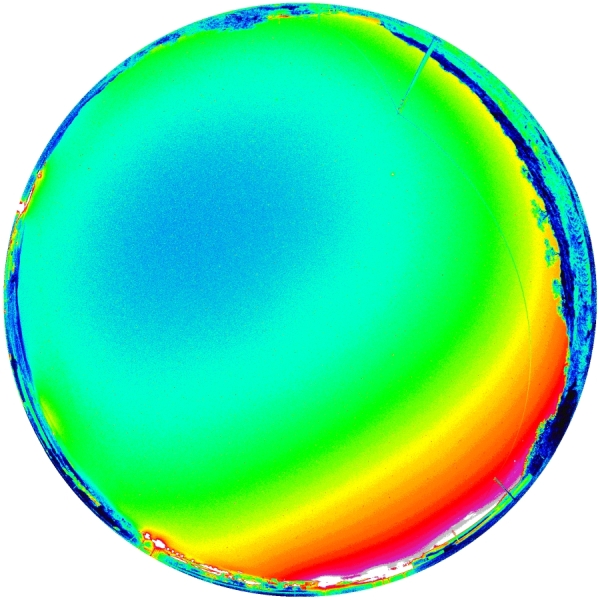}
b) \includegraphics[width=50mm]{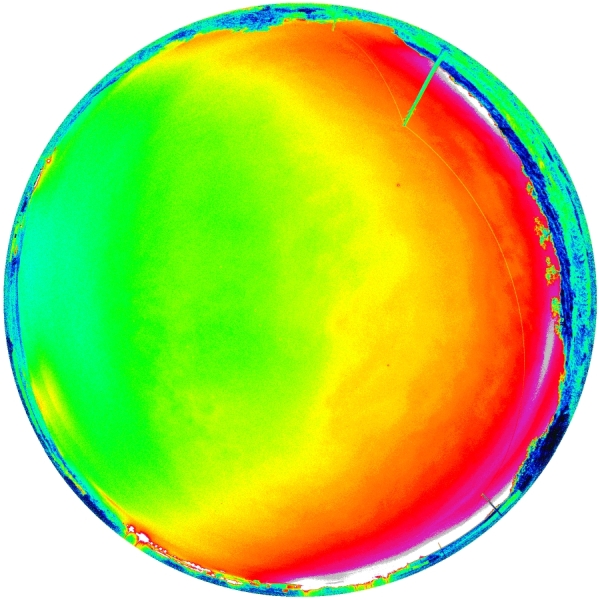}\\
c) \includegraphics[width=50mm]{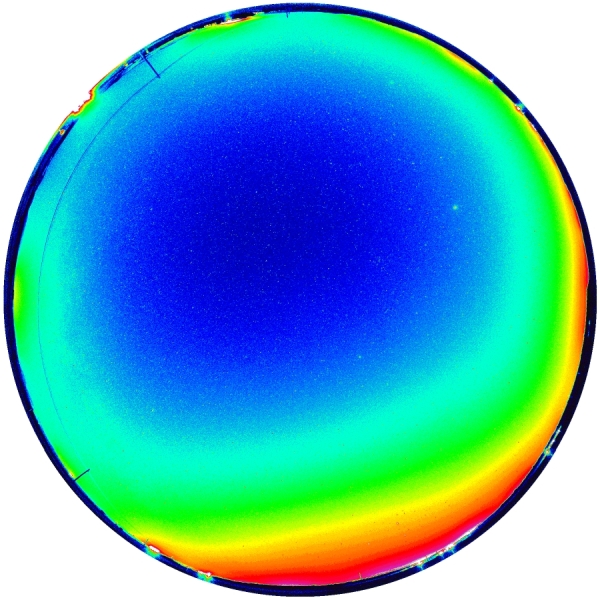}
d) \includegraphics[width=50mm]{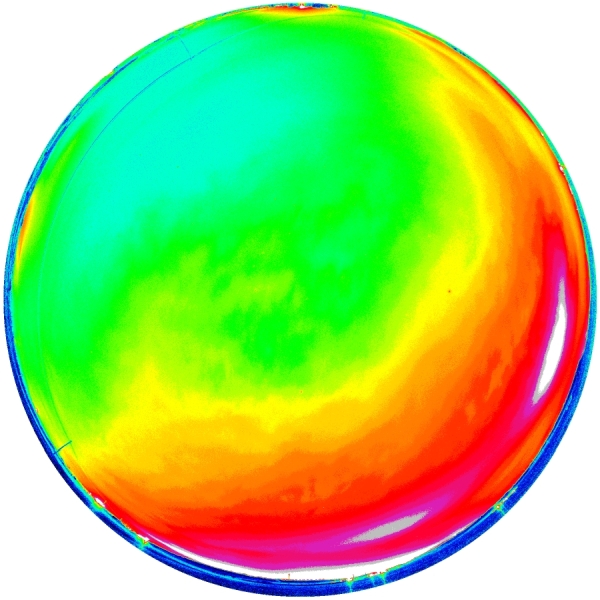}\\
e) \includegraphics[width=50mm]{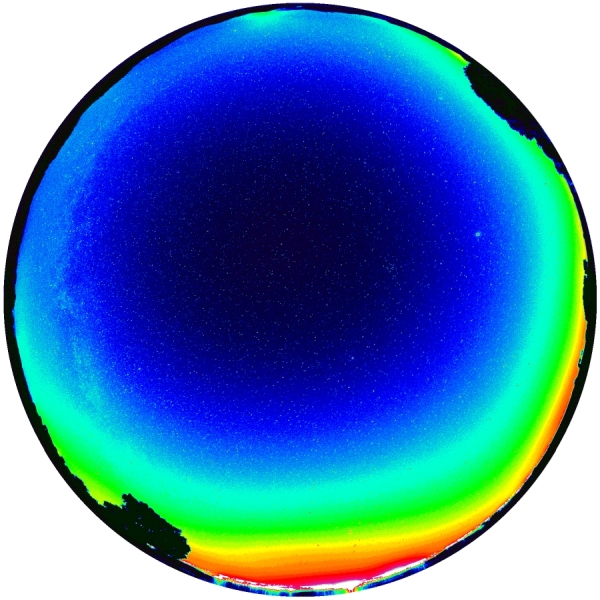}
f) \includegraphics[width=50mm]{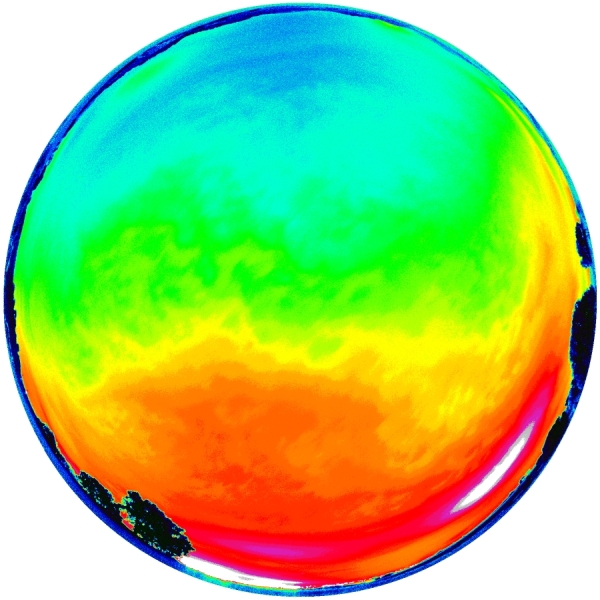}\\
g) \includegraphics[width=50mm]{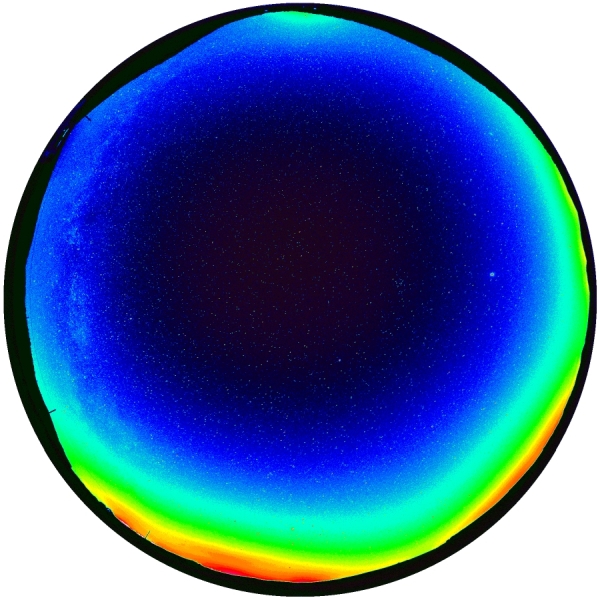}
h) \includegraphics[width=50mm]{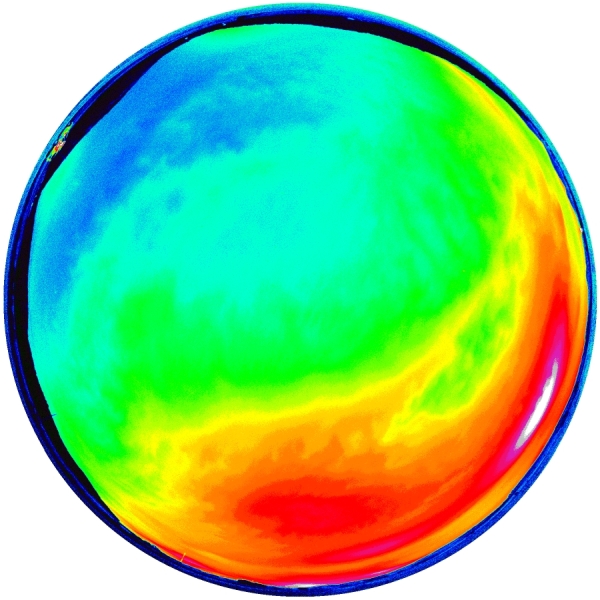}
\caption{All-sky luminance maps along the transect for clear (a, c, e, g) and partly cloudy night (b, d, f, h). The upper row (a, b) was acquired at 1.1 km distance, (c, d) at 2.2 km distance, (e, f) at 4.8 km distance and (g, h) 8.4 km distance to the city center of Balaguer.}
\label{app1}
\end{figure}

\pagebreak
\begin{figure}[h]
\centering
a) \includegraphics[width=50mm]{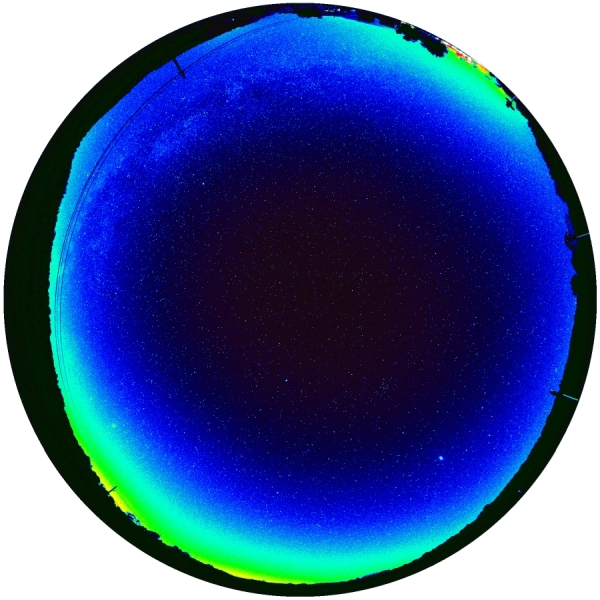}
b) \includegraphics[width=50mm]{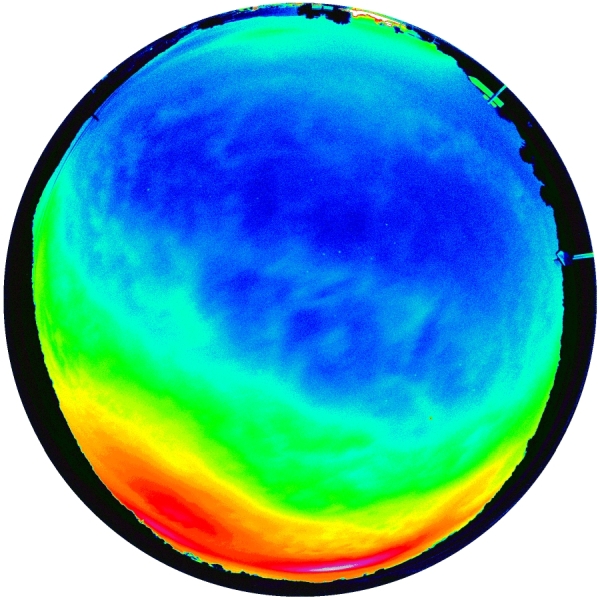}\\
c) \includegraphics[width=50mm]{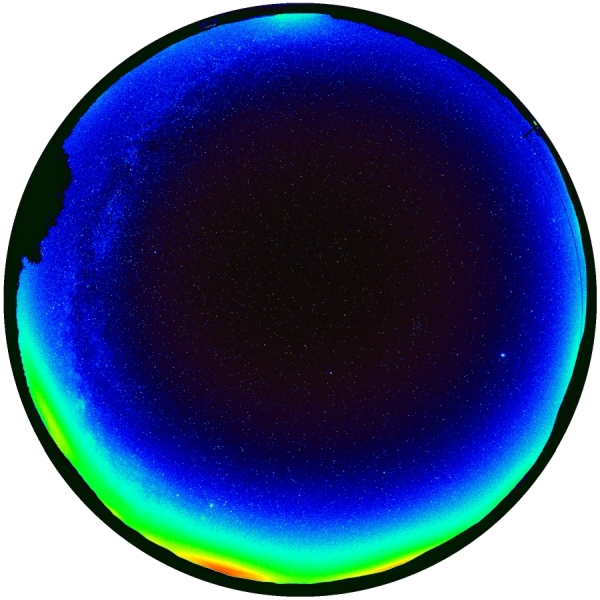}
d) \includegraphics[width=50mm]{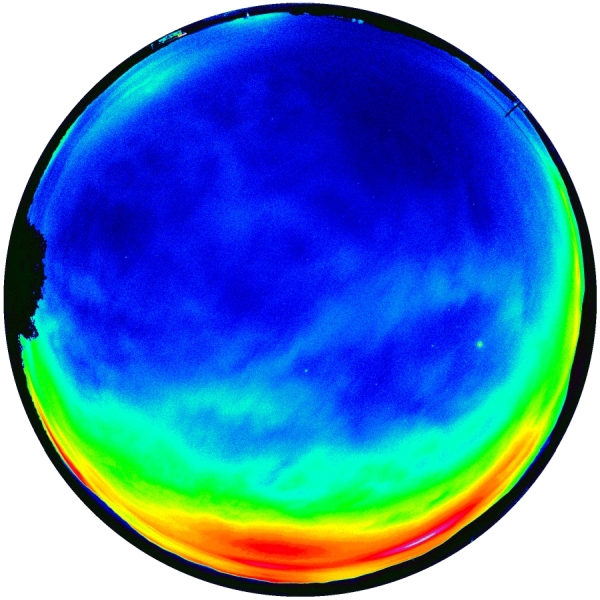}\\
e) \includegraphics[width=50mm]{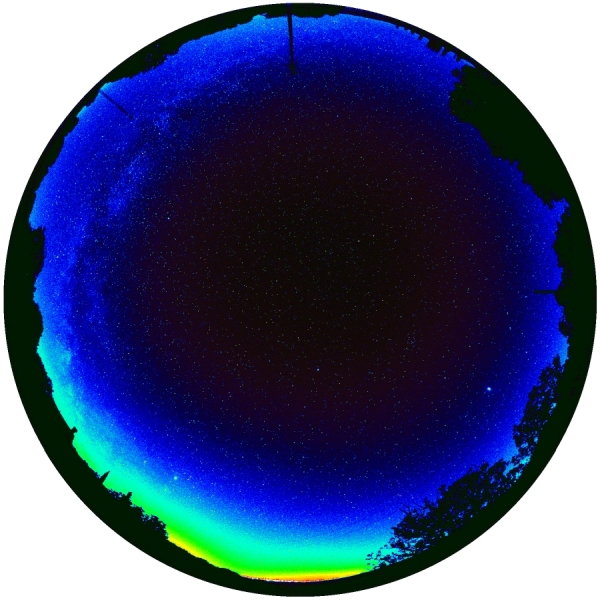}
f) \includegraphics[width=50mm]{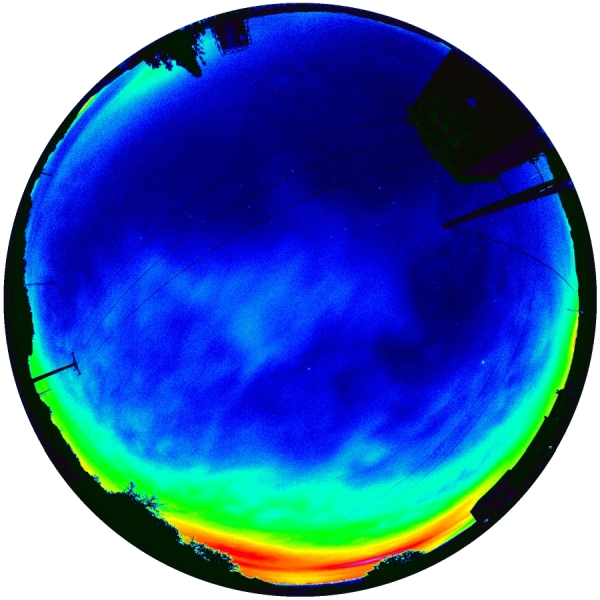}\\
g) \includegraphics[width=50mm]{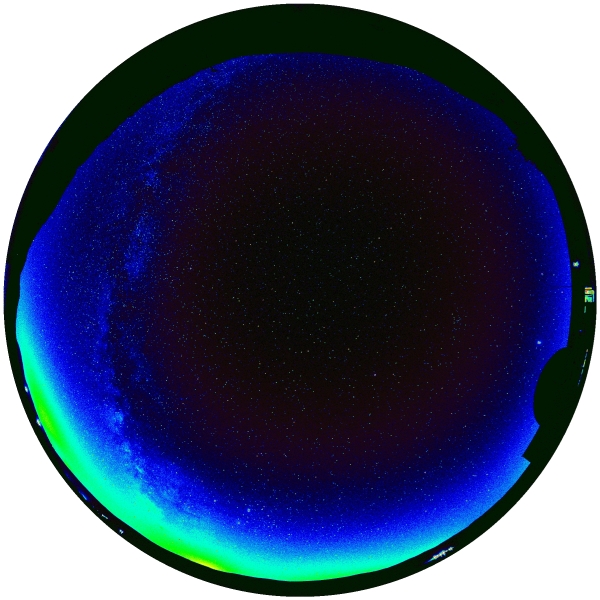}
h) \includegraphics[width=50mm]{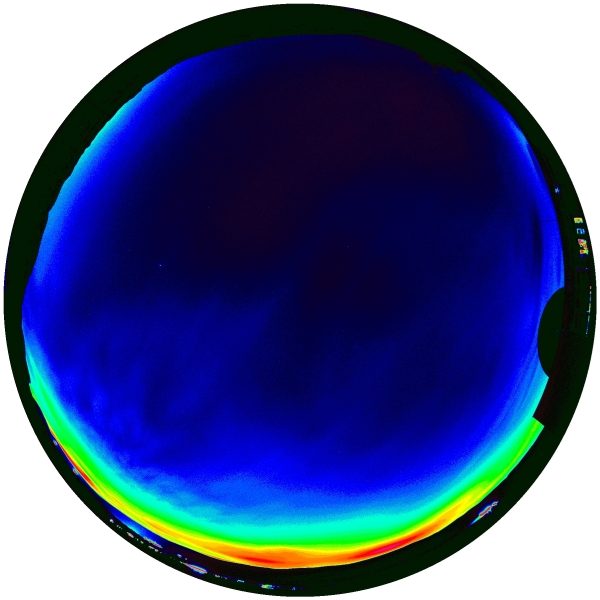}
\caption{All-sky luminance maps along the transect for clear (a, c, e, g) and partly cloudy night (b, d, f, h). The upper row (a, b) was acquired at 13.5 km distance, (c, d) at 17.7 km distance, (e, f) at 22.2 km distance and (g, h) at PAM at 27.2 km distance to the city center of Balaguer.}
\label{app2}
\end{figure}

\end{document}